\documentclass[aip,pof,groupedaddress,showpacs,reprint,english]{revtex4-1}
\usepackage[T1]{fontenc}
\usepackage[utf8]{luainputenc}
\usepackage{textcomp}
\usepackage{amsmath}
\usepackage{amssymb}
\usepackage{graphicx}
\usepackage{color}
\usepackage{ulem}
\makeatletter

\newcommand{\lyxmathsym}[1]{\ifmmode\begingroup\def\b@ld{bold}
  \text{\ifx\math@version\b@ld\bfseries\fi#1}\endgroup\else#1\fi}

\@ifundefined{textcolor}{}
{%
 \definecolor{BLACK}{gray}{0}
 \definecolor{WHITE}{gray}{1}
 \definecolor{RED}{rgb}{1,0,0}
 \definecolor{GREEN}{rgb}{0,0.7,0}
 \definecolor{BLUE}{rgb}{0,0,1}
 \definecolor{CYAN}{cmyk}{1,0,0,0}
 \definecolor{MAGENTA}{cmyk}{0,1,0,0}
 \definecolor{YELLOW}{cmyk}{0,0,1,0}
 }

\usepackage{latexsym}
\newcommand{\W}{7.7cm}

\usepackage{babel}
\begin{document}


\title{Transverse and lateral confinement  effects  on  the oscillations
of a free cylinder in a viscous  flow.}
\author{Luciano Gianorio}
\email{luchogiano@gmail.com}
\affiliation{Grupo de Medios
Porosos, Facultad de Ingenier\'{\i}a, Paseo Colon 850, 1063, Buenos
Aires (Argentina), CONICET (Argentina).}
\author{Maria Veronica D'Angelo}
\email{vdangelo@fi.uba.ar}
\affiliation{Grupo de Medios
Porosos, Facultad de Ingenier\'{\i}a, Paseo Colon 850, 1063, Buenos
Aires (Argentina), CONICET (Argentina).}
\author{Mario Cachile}
\email{mcachil@fi.uba.ar}
\affiliation{Grupo de Medios
Porosos, Facultad de Ingenier\'{\i}a, Paseo Colon 850, 1063, Buenos
Aires (Argentina), CONICET (Argentina).}
\author{Jean-Pierre Hulin}
\email{hulin@fast.u-psud.fr}
\affiliation{Univ Pierre et Marie Curie-Paris 6, Univ Paris-Sud, CNRS, F-91405.
  Lab FAST, B\^at 502, Campus Univ, Orsay, F-91405 (France).}
\author{Harold Auradou}
\email{auradou@fast.u-psud.fr}
\affiliation{Univ Pierre et Marie Curie-Paris 6, Univ Paris-Sud, CNRS, F-91405.
  Lab FAST, B\^at 502, Campus Univ, Orsay, F-91405 (France).}


\date{\today}
\begin{abstract}
The different types of instabilities of free cylinders (diameter $D$, length $L$) have been studied in a  
viscous flow (velocity $U$) between parallel vertical walls of horizontal width $W$ at a distance $H$: the
influence of the confinement parameters $D/H$ and $L/W$ has been investigated.
As $D/H$ increases, there is a transition from stable flow to oscillations transverse to the walls and then to 
a fluttering motion with oscillations of the angle of the axis with respect 
to the horizontal. The two types of oscillations may be superimposed in the transition domain.
The frequency $f$ of the  transverse oscillations is independent of the lateral confinement $L/W$
in the range: $0.055 \le L/W  \le 0.94$ for a given cylinder velocity $V_{cx}$ and increases
 only weakly with $V_{cx}$. These results are accounted for by assuming
 a $2D$  local flow over the cylinder  with a characteristic  velocity independent of $L/W$ for 
 a given $V_{cx}$ value. The experimental values of  $f$ are also independent of the transverse 
 confinement $D/H$. 
 The  frequency $f_f$ of the fluttering motion is significantly lower than $f$: $f_f$ is 
 also nearly  independent of the cylinder diameter and of the flow velocity but decreases significantly as $L/W$
  increases. The fluttering instability is then rather a $3D$ phenomenon involving the full length of the cylinder and
  the clearance between its ends and the side walls.
\end{abstract}
\maketitle
\section{Introduction} \label{intro} 
The transport of elongated particles or microorganisms by a confined flow is
relevant to many industrial applications and natural phenomena. 
This is for instance the case in  bioengineering or  enhanced oil recovery processes or 
in the build-up and structuration of biofilms in flow channels. We are interested in the present work 
in the instabilities of the motion of single elongated particles (here cylinders) free to move in viscous
flows. The characteristic dimensions of the particles are comparable to those of the section of the flow channels
so that confinement  influence very strongly their transport and the occurrence of instabilities.
This study is therefore relevant for instance, to the transport of 
fibers or long bio-particles in micro-fluidic channels or in porous and fractured media.

Previous studies in such flow configurations dealt frequently with the prediction or measurement
of the hydrodynamic forces on static cylinders submitted to a  flow between parallel plates or in 
a rectangular channel~\cite{Richou2004,Richou2005,Semin2009}. 
Investigations of moving  cylinders in such geometries  were often restricted to stable motions~\cite{Faxen1946,Dvinsky87a,Dvinsky87b,Eklund1994,Hu95};  studies of flow instabilities in these geometries
dealt mostly with vortex shedding behind  fixed cylinders 
between parallel walls~\cite{Williamson1996,Camarri2010,Williamson2008}. 
Finally, instabilities occurring during the sedimentation
of different types of objects in a viscous fluid were mostly analysed when no
confinement effects were present~\cite{Ern2012,Assemat2012}.
Periodic fluttering-like motions have also been studied on plates falling in air,
but also without considering the effect of confinement~\cite{Tanabe1994,Belmonte1998,Pesavento2004}.

Regarding  the present configuration, two previous papers  reported experiments and $2D$ numerical simulations of the motion of a tethered~\cite{Semin2011} or
free \cite{Dangelo2013}  horizontal cylinder of diameter $D$ inside a  parallelepiped Hele-Shaw cell where  
a vertical Poiseuille flow of velocity $U$ is established. 
Transverse oscillations of the cylinder in the aperture $H$ of the cell are observed  both  when the cylinder is tethered and when it can move freely  across as well as in the plane of the cell. 
Moreover, in this latter case, the cylinder displays, in addition, oscillations of its rolling angle about its axis and of its vertical position at frequencies respectively equal to and twice that of the transverse oscillations.  
An important feature  is that, in both cases, these oscillations have been observed at  Reynolds numbers $Re $ as low as $20$: this is well below the threshold value generally reported by other authors for vortex shedding in confined geometries \cite{Sahin2004}. This suggests that one deals with a mechanism different from those associated to the destabilization of the wake of fixed bodies. The instability observed here cannot appear if the cylinder is fixed: it  involves likely a feedback effect
originating in the variations of the pressure and viscous forces  induced by the motion of the cylinder. 

In these previous studies the flow was  either  exactly (simulations)~\cite{Semin2011} or approximately (experiments)~\cite{Dangelo2013} two dimensional in the whole cell: in the experiments, the length $L$ of the cylinder was indeed nearly equal to the width $W$ of the cell (experiments) and only cases  in which the cylinder remained horizontal were  studied. The present work studies instead  the influence of the length $L$ ($< W$) on the instabilities; it also deals both with oscillations modes transverse to the cell aperture and with fluttering modes in which the cylinder does not remain horizontal.

  After identifying the different flow regimes,  one studies 
 the influence of the ratio $L/W$ on the instability over the range $0.055 \leq L/W \leq 0.94$: as $L/W$ becomes
 smaller, the influence of the bypass flow between the ends of the 
cylinder and the sides of the cell becomes larger.   Of special interest is the variation of the frequency
$f$  with the  velocities  $V_{cx}$ and $U$ of the cylinder and the flow and the relation between $V_{cx}$ and $U$. 
 Then, the influence of the blockage ratio $D/H$ is investigated over the range of values:  $0.39 \leq D/H \leq 0.77$. One studies finally the fluttering motion of the cylinder in the plane of the cell which appears at large values of $D/H$ and/or $L/W$: the motion of the cylinder displays then periodic variations of the angle of the cylinder with respect to horizontal and oscillatory  displacements parallel to its axis.
\section{Experimental setup and procedure}
\begin{figure}[htbp]
\includegraphics[width=\W]{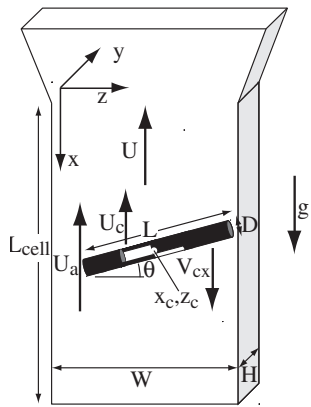} 
\caption{Schematic view of the experimental setup.}
\label{fig:exp} 
\end{figure}
The experimental setup has been described in detail in ref.~\onlinecite{Dangelo2013}.
The length $L_{cell}$,   width $W$ and aperture  $H$ of the Hele-Shaw cell (Fig.~\ref{fig:exp}) are
respectively   equal  to $290$, $90$ and $2.85\, \mathrm{mm}$. 
The flowing fluid is a water-glycerol solution of concentration in weight  $C = 10 \%$.
The viscosity and density of the solution at $T = 25^o C$ are respectively $\mu = 1.153\, \mathrm{mPa.s}$ and
$\rho_f = 1021\, \mathrm{kg/m^3}$.
The flow rate varies  between $0$ and $400 \, \mathrm{ml/mn}$ corresponding to mean velocities 
$-25 \leq U \leq 0 \, \mathrm{mm/s}$ ($U$ is negative for an upward flow velocity since
the vertical axis $U$ is oriented downward). The top part of the cell as a Y-shape so that the local aperture increases
from $2.85$ to $6\, \mathrm{mm}$ over a vertical distance of $48\, \mathrm{mm}$. 
All the experiments are performed using plexiglas cylinders of density $\rho_s = 1.19 \times 10^3 \, \mathrm{kg/m^3}$.
Their lengths range between $5$ and $85 \, \mathrm{mm}$ ($0.055 \le L/W \le 0.94$) and their diameter 
between $1.1$ and $2.2 \, \mathrm{mm}$, ($0.39 \le D/H  \le 0.77$).

At the beginning of the experiment the cylinders are placed horizontally at the top end of the cell and one lets them drift into the
constant aperture region by reducing the flow rate $Q$; $Q$ is then adjusted in order to bring the cylinder at the desired
initial location and is kept constant thereafter during the measurements.

The displacement of the cylinder is monitored by a digital camera viewing the Hele Shaw cell from the front: its resolution is
 $1024 \times 768$ pixels  and the frame rate  $30\, {\rm fps}$.
 In order to analyze the motion of the cylinder, its length is divided into $4$ parts: the two outside ones
 are painted in black while  black staggered stripes parallel to the axis are painted on the  central portions. 
Processing digitally the images provides first the location of the ends of the cylinder by detecting the ends 
of the outer stripes: from  these data one determines then the coordinates $(x_c, z_c)$ of the center of mass 
of the cylinder and its angle $\theta$ with respect to the horizontal. 
As observed previously~\cite{Dangelo2013}, oscillations of the cylinder transverse to the walls of the 
cell are accompanied by  oscillations at the same frequency $f$ of the angle of rotation 
of the cylinder about its axis: the variations of this angle  are estimated by computing
 the transverse displacement of the staggered stripes painted on the cylinder with respect to those of the ends.
The frequency of the oscillations is, here, deduced from the variation of this angle with time.
\section{Different cylinder motion regimes}
\begin{figure}[hbp]
\includegraphics[width=\W]{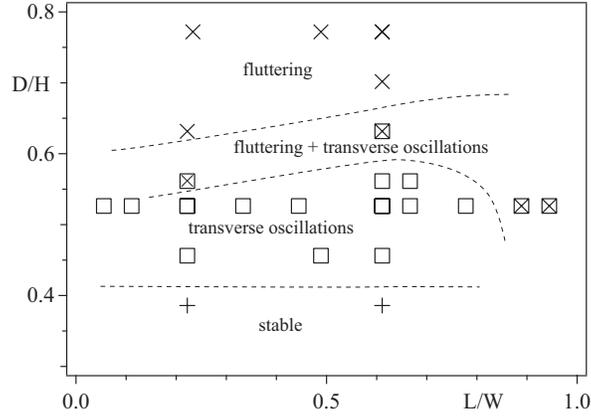} 
\caption{Different types of cylinder motions observed as a function of the ratios $D/H$ and $L/W$
for a plexiglas cylinder and  a water-glycerol solution ($C = 10 \%$): straight trajectory ($+$);
transverse oscillation ($\square$); fluttering+transverse oscillation ($\boxtimes$); fluttering ($\times$).}
\label{fig:regime_map} 
\end{figure}
The different types of motion the free cylinder have been identified for different values of the control parameters
$D$, $L$ and $U$. The diameter and the length of the cylinder were observed to have the largest influence
on the results: we have therefore displayed in Fig.~\ref{fig:regime_map}  a map of the different regimes 
observed as a function of the dimensionless parameters $D/H$ and $L/W$.
\begin{itemize}
\item For  $D/H \lesssim 0.4$, the cylinder
follow a straight stable vertical trajectory with no transverse or side oscillations.
\item  For higher ratios $0.4 \lesssim D/H \lesssim 0.6$, it
displays transverse oscillations. When the length $L$ becomes of the order of $W$ ($L/W \gtrsim 0.9$), 
a fluttering motion   is superimposed onto the transverse oscillations ($D/H = 0.53$): it corresponds to a periodic variation
of the angle $\theta$  with the horizontal with a frequency  significantly lower than that of the transverse
oscillations.
\item For $D/H \gtrsim 0.6$ a fluttering motion without transverse oscillation generally occurs
except for $L/W = 0.61$, $D/H = 0.63$ in which case the two types of oscillations are again superimposed.
\end{itemize}
In short, increasing the ratio $D/H$ and, therefore, the transverse confinement results
in a transition from stable flow to transverse oscillations and then to a fluttering motion:
moreover, fluttering appears earlier for strong longitudinal confinements.
\section{Influence of the  confinement on the   transverse oscillations}
\subsection{Influence of the cylinder length}
\begin{figure}[htbp]
\includegraphics[width=\W]{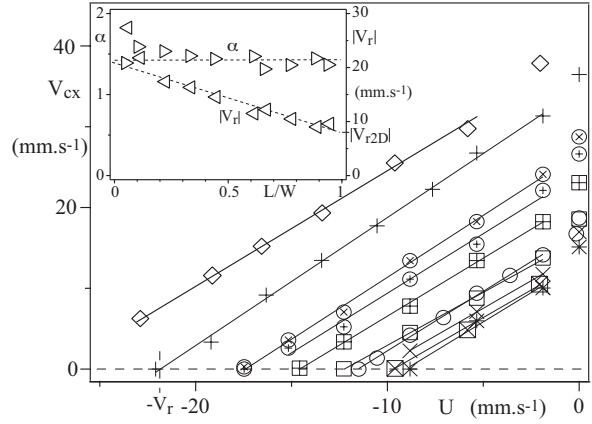} 
\caption{Influence of the length $L$ of plexiglas cylinders of constant diameter $D = 1.5\, \mathrm{mm}$ ($D/H = 0.53$) on
 the variation of  cylinder velocity $V_{cx}$ with the  velocity $U$ of a flow of water  with $10\%$ glycerol.
  Symbols:  experimental data; straight lines: linear regressions 
over these data (excluding point $U = 0$). 
$L/W = 0.055$ ($\lozenge$), $0.11$ ($+$), $0.22$ ($\otimes$), $0.33$ ($\oplus$), $0.44$ ($\boxplus$), $0.61$ ($\bigcirc$), $0.67$ ($\Box$), $0.77$ ($\times$),  $0.89$ ($\ast$) and $0.94$ ($\boxtimes$).
Inset: variation as a function of the cylinder length $L$ of the  slope $\alpha$ of the linear regressions ($\triangleright$)
and of the  velocity $|V_r|$ ($\triangleleft$).}
\label{fig:Vc_U_L} 
\end{figure}
The influence of the lateral confinement parameter $L/W$ on the transverse 
oscillations has first been studied: experiments have been performed   for free  cylinders of  diameter 
$D = 1.5\, \mathrm{mm}$ ($D/H = 0.53$) 
and  $L$ varying between $5$ and $85 \, \mathrm{mm}$ ($0.055 \le L/W \le 0.94$).

A first important characteristic is the variation of the velocity $V_{cx}$ of the cylinder as a function 
of that of the flow ($U$)  which is here oriented upward and, therefore, negative. 
The main graph in Fig.~\ref{fig:Vc_U_L} shows
that $V_{cx}$ varies  linearly with $U$ (data points corresponding to $U = 0$
are however above the linear trend).
For the  curves of  Fig.~\ref{fig:Vc_U_L} corresponding to $L/W \le 0.77$,  only  
transverse oscillations occur:  the axis of the cylinder remains horizontal and no flutter is visible. 
For $L/W = 0.89$ ($\ast$) and $L/W = 0.94$ ($\boxtimes$), the cylinder both flutters and oscillates transversally. 

The straight lines on the main graph of Fig.~\ref{fig:Vc_U_L} correspond to  a linear regression on the data
according to the equation:
\begin{equation}
 V_{cx} = \alpha (U - V_r);
 \label{eq:Vr}
 \end{equation}
The variations with $L/W$ of the slope $\alpha = \mathrm{d}V_{cx}/\mathrm{d}U$ of the regression lines and
 of $V_r$ are plotted in the inset:
 $\alpha$  depends only weakly  on  $L$, even when fluttering occurs   ($\triangleright$ symbols) and  
its values are all in the range $1.4 \pm 0.1$.
From Eq:\ref{eq:Vr},   $V_r\, (< 0)$  is the velocity of the upward flow
at  which the cylinder remains at a constant average vertical position~\cite{Dangelo2013};
more generally $V_r$ can be considered as a relative velocity of the fluid and the cylinder: the fact
that it remains constant for a free cylinder as $V_{cx}$ suggests that the drag force on the 
cylinder is determined by $V_r$ and remains constant as $V_{cx}$ varies for a given free cylinder
in order to balance its weight.
In contrast to  $\alpha$, $|V_r|$ decreases   as $L$ increases  ($\triangleleft$ symbols):
 the limit  of $V_r$ as $L \rightarrow W$  corresponds to the value  $V_{r2D}$ for a $2D$
 configuration with, here: $V_{r2D} = -9\, \mathrm{mm.s^{-1}}$. The slope of the variation 
 of $|V_r|$ with $L/W$ is almost constant except at the lowest values for which it increases
 sharply.

\begin{figure}[htbp]
\includegraphics[width=\W]{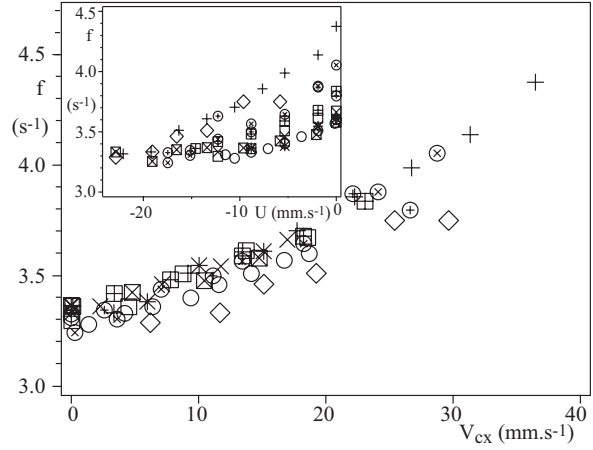} 
\caption{Experimental variation of the transverse oscillation frequency $f$ as a function of the
 cylinder velocity $V_{cx}$
for plexiglas cylinders of  different dimensionless lengths $L/W$ and constant dimensionless
diameter $D/H = 0.53$.
Inset : same frequency data as in the main graph plotted as a function of the mean flow velocity $U$. 
In both graphs, the  symbols are  the same as in Fig.~\ref{fig:Vc_U_L}.}
\label{fig:f_U_L} 
\end{figure}
Fig.~\ref{fig:f_U_L}, displays the variation of  the frequency  $f$ with  the cylinder velocity $V_{cx}$ 
for the different  lengths $L$: one observes then an excellent collapse of the different curves onto a common 
weakly increasing trend, with, for $V_{cx} = 0$, a frequency  $f = 3.3 \pm 0.1 \, \mathrm{Hz}$. The data points
are much more dispersed  when $f$ is plotted as a function of $U$ (see inset). 
 We explain now    the above result, namely that, for free cylinders,  the frequency $f$ is independent of $L/W$ 
 when the velocity $V_{cx}$ is  kept constant. 
  When  $L \rightarrow W$ (like in Ref.~\onlinecite{Dangelo2013}), flow is  
 two dimensional with a zero bypass flow between the ends of the cylinder and the side walls. 
 The balance, per unit length, between the weight of the cylinder and the vertical hydrodynamic
force  is then~\cite{Semin2009}:  
\begin{equation}
 \lambda_{p2D}\, \mu\, {U}  - \lambda_{s2D}\, \mu\, {V}_{cx} = - (\rho_s - \rho_f) A\,  g
\label{eq:drag}
\end{equation}
 ($A$ is the cylinder section).  Eq.~(\ref{eq:drag}) can then be rewritten in the form similar
 to Eq.~(\ref{eq:Vr}):
\begin{equation}
{V}_{cx} = \frac{\lambda_{p2D}}{\lambda_{s2D}} U + \frac{(\rho_s - \rho_f) A g}{\lambda_{s2D}\, \mu} = \alpha_{2D} (U - V_{r2D}), \label{eq:Vr_2D}
\end{equation} 
in which $V_{r2D}$ and $\alpha_{2D}$ are constant with $U$.
$V_{r2D}$ and $\alpha_{2D}$ will be equal to the limits of $V_r$ and $\alpha$ when $L/W \rightarrow 1$
with, for $D/H = 0.53$, from the inset of Fig.~\ref{fig:Vc_U_L}: $|V_{r2D}| = 9\, \mathrm{mm.s^{-1}}$ and $\alpha_{2D} = 1.4$
 
In the general case  $L < W$, the local flow {in the part of the aperture occupied by}  the cylinder is  still  assumed
 to be  two dimensional: more precisely, one assumes that the velocity component $v_z$ is negligible and that 
 $v_x(x,y)$ and $v_y(x,y)$ are independent of $z$ along the length $L$ of the cylinder.
The results to be discussed below suggest that this assumption is valid for $L/D \gtrsim 7$.

The balance of forces per unit length on the cylinder should then be  the same as for $L = W$:
Eqs.~\ref{eq:drag} and ~\ref{eq:Vr_2D} remain then  valid  with  the  same 
parameters  $\lambda_{s2D}$ and $\lambda_{p2D}$  (or $V_{r2D}$ and $\alpha_{2D}$) 
provided  $U$ is replaced by  a local velocity $U_{loc}$ constant along the length $L$. Then:
\begin{equation}
{V}_{cx} = \frac{\lambda_{p2D}}{\lambda_{s2D}} U_{loc} + \frac{(\rho_s - \rho_f) A g}{\lambda_{s2D}\, \mu} =
 \alpha_{2D}(U_{loc} - V_{r2D}). 
\label{eq:Vr_loc}
\end{equation} 
$U_{loc}$ is related to  $V_{cx}$ and $V_{r2D}$ by Eq.~(\ref{eq:Vr_loc}) and is, therefore, 
 independent of $L/W$. Since the velocities $U_{loc}$ and $V_{cx}$ determine completely
the local flow on the  cylinder and, therefore, the frequency $f$, the latter will also  be independent of $L/W$:
this explains the excellent coincidence of the curves of Fig.~\ref{fig:f_U_L}.

In order to understand the relation between  $V_r$ and  $L/W$ displayed in the inset of 
Fig.~\ref{fig:Vc_U_L}, one estimates first the difference $U_{loc} - U$. 
The flow in each clearance of  width $(L - W)/2$ between the ends of the cylinder and the sides of the cell (Fig.~\ref{fig:exp}) is, like  around the cylinder, assumed to be viscous and  two dimensional:   the corresponding velocity 
$U_a$ averaged over the aperture $H$ is then  constant with $z$ along its width $W - L$. 
Applying mass conservation, $U$, $U_a$ and $U_{loc}$  satisfy $U = (U_{loc} \, L + U_a \,(W - L))/W$.
Moreover, the constant value of $\alpha$, in particular as $L \rightarrow W$  allows one to take $\alpha = \alpha_{2D}$.
Combining  these two latter  results with  Eqs.~(\ref{eq:Vr}) and (\ref{eq:Vr_loc}) leads then to:
\begin{equation}
V_r - V_{r2D} =  U - U_{loc} = \frac{W - L}{W} (U_a - U_{loc})
\label{Vr_Vr2D}
\end{equation}

Taking for simplicity $V_{cx} = 0$, momentum conservation requires that  the force due to the  pressure drop 
$\Delta p$ between the  upstream and downstream sides of the cylinder balances its effective weight per unit length. Then:
$\Delta p\, H = (\rho_s - \rho_f) A  g$ so that $\Delta p$ is independent of $L/W$. Assuming a transverse pressure
equilibrium, the pressure drop   across  the clearance between the cylinder and the walls must also be $\Delta p$. 
Under the above assumptions of a  $2D$ viscous flow, the  velocity $U_a$ is proportional
to $\Delta p$ with a coefficient independent of the width $W - L$. Like $\Delta p$, $U_a$ is then constant with $L/W$ 
and Eq.~(\ref{Vr_Vr2D}) predicts   the linear variation of $V_r$ with $L/W$  observed experimentally. 
 Still for $V_{cx} = 0$ and $D/H = 0.53$ one has, from Eq.~(\ref{eq:Vr_loc}):  
 $U_{loc} = V_{r2D}   = -9\, \mathrm{mm.s^{-1}}$ (see above for the determination of $V_{r2D}$). 
 Taking $L = 0$ in Eq.~(\ref{Vr_Vr2D})  leads then to:  $U_a = V_r(L/W \rightarrow 0) = -27\, \mathrm{mm.s^{-1}}$.
\subsection{Influence of the  diameter on the transverse oscillations}\label{influ_D}
\begin{figure}[h!]
\includegraphics[width=\W]{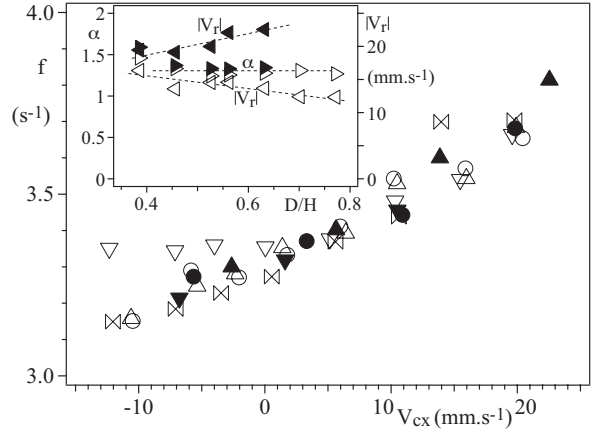} 
\caption{Experimental variation of the transverse oscillation frequency as a function of the mean flow velocity $U$
for plexiglas cylinders of different diameter to aperture ratios:  $D/H = 0.46$ ($\triangledown$, $\blacktriangledown$), 
$0.53$ ($\bigcirc$, $\bullet$), $0.56$ ($\triangle$, $\blacktriangle$),  $0.63$ ($\Join$). 
Open symbols: $L/W = 0.61$; black symbols: $L/W = 0.22$.
Flowing fluid: water-glycerol solution ($C = 10\,\%$). Inset: variation  of the  slope $\alpha$ and
 the velocity $V_r$ with the diameter $D$. 
Data points corresponding to $D/W = 0.39$ (stable regime), $0.7$ and $0.77$ (pure fluttering regime)  have been
added for comparison.}
\label{fig:f_D_U} 
\end{figure}
The influence of the transverse confinement parameter $D/H$ on the transverse oscillations has been investigated
by  using several  cylinders with different diameters and for two different 
lateral confinements ($L/W = 0.61$ and $L/W = 0.22$): the  values of $D/H$ belonged
to the interval $0.39 \le D/H \le 0.77$. Transverse oscillations were observed  in the  range $0.46 \le D/H \le 0.63$. 

The inset of Fig.~\ref{fig:f_D_U} displays the variation of the parameters $\alpha$ and $V_r$ with $D/H$: data   points
corresponding to  pure  fluttering  ($D/H = 0.7$ and $0.77$) or to  stable   ($D/H = 0.39$) regimes have also been
 included in this graph. The experimental value of $\alpha$ is independent of $D/H$ with 
  $\alpha = 1.34 \pm 0.1$ for both transverse confinement ratios $L_c/W$;
moreover, the transition to the stable or fluttering regimes does not result in any variation of $\alpha$.

The velocity $V_r$ decreases smoothly by $30\%$ as $D/H$ varies from $0.39$
to $0.77$ for $L/W = 0.61$ and increases by $15\%$ as $D/H$ varies from $0.39$
to $0.63$ for $L/W = 0.22$: like for $\alpha$, there is no visible influence of the transition from a flow regime
to another.

The variation of the  oscillation frequency with the cylinder velocity $V_{cx}$ is plotted in the main
graph of Fig.~\ref{fig:f_D_U} for these same cylinders. All  data points correspond  to pure transverse oscillations  except for $L/W = 0.61$ and $D/H = 0.63$ (Fig.~\ref{fig:regime_map}): in this latter  case, transverse oscillations  and fluttering occur simultaneously.  
The frequency $f$ is also  remarkably independent of the ratio $D/H$ for all values of $D/H$,
 except for the smallest diameter  $D/H = 0.46$ and for  $L/W = 0.61$: in this   particular case, the common trend 
 of variation of $f$ with $V_{cx}$ is only followed for  $V_{cx} > 0$. but the values of $f$ are higher for $V_{cx} \le 0$.
No special feature of the variations is observable when fluttering is superimposed onto transverse oscillations.
The curves corresponding to the two different values of $L/W$ ($0.22$ and $0.61$) also coincide which generalizes the results 
obtained for $D/H = 0.53$ and displayed in Fig.~\ref{fig:f_U_L}. 
\section{Fluttering oscillations of the cylinder}
\begin{figure}[htbp]
\includegraphics[width=13 cm]{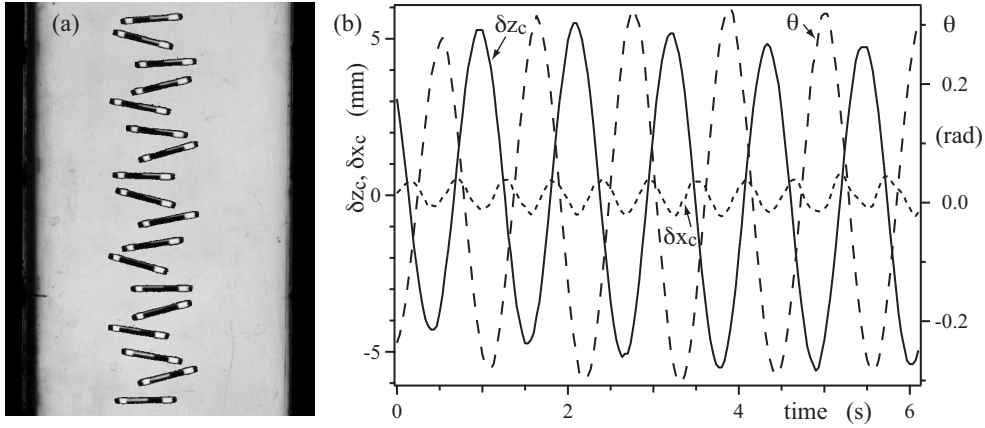} 
\caption{a) Successive views of the cylinder taken at  time intervals $\Delta t = 1/3\, \mathrm{s}$ in the fluttering regime. 
b) Variations as a function of time in the same experiment of the geometrical parameters characterizing the
motion of the cylinder in the fluttering regime;  $\delta z_c$: distance from the vertical axis of symmetry of the cell (continuous line), $\delta x_c$:
deviation of the vertical coordinate from a linear variation with time (dotted  line), $\theta$: angle of the axis with respect to the 
horizontal (dashed line).  $L/W = 0.22$, $D/H = 0.63$, $U = 6.6\, \mathrm{mm.s^1}$.}
\label{fig:flutter_t} 
\end{figure}
The fluttering instability is characterized by oscillations of the angle $\theta$ of the axis of the cylinder with respect to
the horizontal (Fig.~\ref{fig:flutter_t}a) and dashed line in Fig.~\ref{fig:flutter_t}b. These angular oscillations are accompanied by synchronous variations of the lateral displacement $\delta z_c$ of the center of mass  (continuous line):  the angle $|\theta|$ reaches an extremal value  shortly after the end of the cylinder is closest to one of the sides of the cell.

The fluttering motion also induces fluctuations of the vertical velocity $v_x$ of the cylinder. These variations  are 
visualized in the figure (dotted line) from the deviation $\delta x_c$ of the vertical coordinate from the linear trend which would 
correspond to a constant velocity: $\delta x_c$ oscillates at twice the fluttering frequency indicating that negative and positive deviations of the angle $\theta$ have the same influence on the velocity. In the transverse oscillations regime, 
vertical oscillations at a frequency $2f$ have also been observed although, in this case, the cylinder 
remained horizontal~\cite{Dangelo2013}.

\begin{figure}[htbp]
\includegraphics[width=\W]{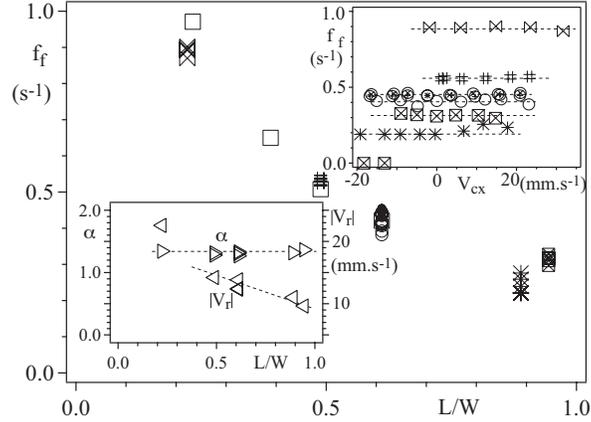} 
\caption{Experimental variation of the fluttering frequency $f_f$ for a water-glycerol solution   ($C = 10\,\%$)
as a function of the normalised length $L/W$ for plexiglas cylinders. Inset at lower left: variation of the  slope $\alpha$ ($\vartriangleright$) and   the velocity $V_r$ ($\vartriangleleft$)  with $L/W$ for different diameters ($0.53 \le D/H \le 0.77$). 
Inset at upper right: variation of $f_f$ with $V_{cx}$ for  cylinders
with different values of $D/H$ and $L/W$ ($f = 0$ means: no oscillation).
Dimensionless length and diameter  of the cylinders: $D/H = 0.63$, $L/W = 0.22$ ($\Join$); 
$D/H = 0.77$, $0.22 \le L/W \le 0.61$ ($\square$);  $D/H = 0.77$, $L/W = 0.49$ ($\#$); $D/H = 0.53$, $L/W = 0.89$ ($\ast$);
$D/H = 0.53$, $L/W = 0.94$ ($\boxtimes$); $D/H = 0.63$, $L/W = 0.61$ ($\bigcirc$); $D/H = 0.7$, $L/W = 0.61$ ($\oplus$);$D/H = 0.77$, $L/W = 0.61$ ($\otimes$)}
\label{fig:ff_Lc_Vr} 
\end{figure}
Fig.~\ref{fig:ff_Lc_Vr} displays  variations of the fluttering frequency $f_f$ as a function of the velocity $V_{cx}$ or
the lateral confinement $L/W$ for different pairs of values of the ratios  $L/W$ and $D/H$.
As mentioned above, the fluttering instability  is  observed  for large values of $D/H \ge 0.63$ either 
alone or superimposed onto transverse oscillations (see Fig.~\ref{fig:regime_map}). For 
$D/H = 0.53$, fluttering is only observed (together with transverse oscillations) for the largest ratios $L/W \ge 0.89$.

A first specific feature of this instability is that the frequency $f_f$ is more than  $3$ times lower than that
 of the transverse oscillations; $f_f$ decreases  significantly with $L/W$, {\it e.g.} by a
 factor $3$ as $L/W$ increases from $0.22$ to $0.9$ (main graph of Fig.~\ref{fig:ff_Lc_Vr}).

The strong influence of $L/W$ on $f_f$ suggests that these oscillations
are driven by the dissymmetry  between the bypass flows  at the two ends of the cylinders
when it moves laterally  ($\delta z_c \neq 0$):
the forces at the two ends of the cylinder are then unequal, which creates a torque 
 that rotates it and a  lateral force inducing a sideways motion. 

Finally, for a given cylinder, $f_f$ is independent of the velocity $V_{cx}$ (and on $U$, too) 
as can be seen in the upper inset of Fig.~\ref{fig:ff_Lc_Vr}. 
In this same graph, ones observes that the frequencies  $f_f$ corresponding to the same ratio
 $L/W = 0.61$ and to different values of $D/H$ ($0.63$, $0.7$, $0.77$) coincide at all velocities
  (($\bigcirc$), ($\oplus$ and ($\otimes$) symbols). Similarly, in the main graph, for $L/W = 0.22$, the
   values of  $f_f$ corresponding to $D/H = 0.63$ and $0.77$ are   nearly equal.

Regarding the mean vertical velocity $V_{cx}$, the slope $\alpha$ of the variation with $U$  is  practically
independent of $L/W$  (insert at lower left of Fig.~\ref{fig:Vc_U_L}): the common value is the same as that found
previously in the stable and transverse oscillation regimes  (insert of Fig.~\ref{fig:Vc_U_L}) Also, like in the 
case of transverse oscillations, the velocity  $|V_r|$ decreases significantly as the ratio $L/W$ increases;
the value of $V_r$ is also nearly independent of $D/H$. 
 
 At a first glance, these fluttering instabilities have visual similarities with those observed 
 for falling  sheets (or leaves)~\cite{Tanabe1994,Belmonte1998,Pesavento2004}:  these latter 
 experiments are however realized in unconfined configurations. These instabilities, which take 
 place in unconfined configurations, take however place at larger Reynolds numbers: they involve 
 vortex shedding from the edges of the sheets in contrast with the present ones.
\section{Conclusion}
The present experiments demonstrate that the motion of a buoyant cylinder in a vertical viscous Hele Shaw cell flow may display 
oscillatory transverse and/or fluttering instabilities depending   on the value of the two
 confinement parameters $L/W$ and $D/H$.  
For  $0.2 \le L/W  \le 0.8$, for instance,  one shifts continuously  from the  stable regime
 to the transverse oscillations and then to  fluttering as $D/H$ increases; at the transition, 
 the  two oscillatory instabilities  may, in addition, be superimposed. These instabilities are controled by the relative velocity between the fiber and the fluid. They are observed for Reynolds numbers (based on the relative velocity) as low as $20$: the mechanisms of the instabilities are thus different from those associated to destabilization of the wake at the rear of a fiber.

In an approximate description, the  transverse instability  is considered as a  $2D$ one, corresponding 
to a local relative  velocity  with the same value $V_{r2D}$ as for a cylinder of length $L=W$ (the transverse deviations of
the flow lines are neglected). $V_{r2D}$ is determined by the cylinder velocity $V_{cx}$ and a local flow velocity $U_{loc}$.
Both  $V_{r2D}$ and $U_{loc}$
 cannot be determined directly with the present setup: they can however be assumed to be equal respectively to the 
 experimental values of $V_{r}$ and $U$ in the limiting case $L/W \rightarrow 1$.
As a result, the frequency $f$ is independent of $L/W$ for a constant velocity $V_{cx}$ (but depends instead 
on $L/W$ for a constant velocity $U$); also, $f$ increases by less than $15  \%$  when $V_{cx}$ varies from 
$0$ to $20\, \mathrm{mm.s^{-1}}$ in agreement with the results report in \cite{Dangelo2013}.
This $2D$  description is not valid for the shortest cylinders ($L/W = 0.055$) of aspect ratio 
$L/D = 1.75$. 

The above discussion is only valid for free cylinders.
For tethered ones~\cite{Semin2011} for which $V_{cx} = 0$,
 there is no longer an equilibrium between the hydrodynamic forces on the cylinder and its weight because of
 the tension of the supporting threads.  In this case,  the frequency $f$ depends both on  $U$ 
 and on $L/W$:  $f$ is indeed determined in this case by the local velocity $U_{loc}$. If the fiber and the fluid have the same density (this situation was considered by Berthet~\cite{Berthet2013}), we have $V_{cx}=\alpha U$, and a relative velocity $V_r=0$. In this case, fluttering and oscillations in the gap will not be observed.  
 
Reverting to the free case, the frequency  $f(V_{cx})$ is also found experimentally to be  independent 
of the dimensionless diameter  $D/H$: this result is quite surprising since,  a  first view, several mechanisms
might  induce a variation of $f$. Due to the  lack of dependence of $f$ on $L/W$, one can consider 
this problem for simplicity in the  $2D$ case equivalent  to  $L/W =1$.  
First, increasing $D/H$ increases the section and, therefore, the mass of the cylinder which should reduce
the frequency $f$.  Increasing $D$ also 
 reduces the clearance $H - D$ and enhances the velocity (and Bernoulli pressure) variations: this should
 instead increase the value of $f$.
 Varying $D$  also influences the relative velocity $V_{r2D}$ and, as a result: $f$. Increasing $D$ 
 increases first the weight of the cylinder which is the driving force in Eq.~(\ref{eq:drag}); it should
  also increase the drag by reducing the clearance between the cylinder and the front walls. These two 
 effects will respectively tend to increase and reduce the relative velocity (and therefore the frequency). $2D$ 
 numerical simulations should allow one to determine the relative magnitude of the different effects and
 whether they compensate each other. 

In contrast, the fluttering instability is  strongly related  to  the variations of the distances between 
the ends of the cylinder and the sides of the cell: understanding it  requires a model at the scale
of the full width $W$ of the Hele Shaw cell.

In spite of these differences, the transverse and fluttering instabilities of free cylinders share several 
common properties. Both  $f$ and $f_f$ depend weakly, or not at all, on the 
 velocity $V_{cx}$ for a given cylinder:  in both cases, this results from the fact 
 that, as mentioned above, these frequencies are determined mainly by a relative velocity of the cylinder 
 and the fluid: the latter remains constant with the flow velocity, again in order to keep the balance 
 between the hydrodynamic  forces and the weight of the cylinder. 
   Also,  $f$ and $f_f$ are both independent of the cylinder diameter: this result involves likely 
   a compensation between different effects and will require $2D$ numerical 
   simulations in  order to be explained. 

Further studies are needed to understand better the analogies and  differences of these two types 
of instabilities as well as the weak dependence of variables like $\alpha$ on all the control parameters 
 investigated ($D/H$, $L/W$). 
Regarding the fluttering instability, the characteristics of the fluid are important parameters
to be investigated. 
\begin{acknowledgments} 
We thank 
B. Semin for his careful reading of the manuscript and his useful comments and 
J.E. Wesfreid for useful suggestions. 
We acknowledge the RTRA Triangle de la Physique
and the LIA PMF-FMF (Franco-Argentinian International Associated Laboratory
in the Physics and Mechanics of Fluids). The work of one of us (VD) was supported by
a Bernardo Houssay grant allocated by the Argentinian and French
ministries of research. 
\end{acknowledgments} 

\end{document}